# Cross Feshbach resonance

M. Navadeh-Toupchi , N. Takemura, M. D. Anderson, D. Y. Oberli, M. T. Portella-Oberli

*Institute of Physics, School of Basic Sciences, Ecole Polytechnique Fédérale de Lausanne, CH-1015, Lausanne, Switzerland*
(Dated: March, 2018)

**Feshbach resonance occurs when a pair of free particles is resonantly coupled to a molecular bound state. In the field of ultracold quantum gases, atomic Feshbach resonances [1, 2] became a usual tool for tailoring atomic interactions opening up many new applications in this field [3-8]. In a semiconductor microcavity, the Feshbach resonance appears when two lower polaritons are coupled to the molecular biexciton state [9]. Here, we demonstrate the existence of a cross Feshbach resonance for which a pair of polaritons, lower together with upper, effectively couples to the biexciton state. This demonstration is a crucial step towards the efficient generation of entangled photon pairs in a semiconductor microcavity [10, 11]. The existence of a Cross Feshbach resonance establishes the condition to convert a pair of upper and lower polaritons via the biexciton state into two lower polaritons, paving the way for the generation of momentum and polarization entangled photons.**

A Feshbach resonance appears whenever two free particles resonantly couple to a molecular bound state. Near the resonance, the strength of the interaction between the particles is modified and its sign changes at resonance. Since the demonstration of atomic Feshbach resonances [1,2], they have been extensively used to control the interactions in atomic Bose Einstein condensates by tuning their strength within the condensate [3,4]. As key examples, let us cite the investigation of the explosion and instability of Bose Einstein condensates in an attractively interacting Bose gas [5] and the observation of the BEC-BCS crossover in degenerate fermionic cold atomic gases when tuning the atomic interactions from repulsive to attractive [6]. Also, the Bogoliubov theory of quantum depletion has been recently verified by tuning the interaction strength in a condensate by means of the Feshbach resonance [8]. Moreover, thanks to the Feshbach resonance, gases of ultracold atoms provide a platform for exploring models of many-body physics in condensed matter systems. Recently, the d-wave interactions for simulating high temperature superconductivity have been implemented [8]. The atomic systems with a Feshbach resonance share a common characteristic: they feature tunable repulsive or attractive interaction strength. This fascinating property may also be implemented directly with a system of polaritons in semiconductor optical microcavities. A polaritonic Feshbach resonance has been demonstrated when two lower polaritons are efficiently coupled to the biexciton, which is the quasiparticle analogue of the molecular state in a semiconductor system [9].

In a semiconductor microcavity, the strong coupling between excitons and photons gives rise to two new eigenstates the lower and the upper polaritons [12]. The hybrid matter-light nature of these quasiparticles provides both nonlinear behavior due to excitonic interactions and coherent properties because of their photonic component. By exploring these interacting polariton systems, a wide range of research fields have emerged, such as polariton condensates [13, 14], quantum fluids [15,16], quantized vortices [17], Dirac cones on honeycomb lattice of polaritons [18], the observation of spontaneous spin bifurcations in a spinor polariton gas [19], and polariton squeezing [20,21].

There is a direct correspondence between the spin of the exciton and the polarization of the cavity photon, thus the spin of a polariton can be matched to its photonic part, being either by right or left circular polarized. The spinor character of spinor polariton interactions [22-24] offers a wide range of physics to explore [25-27]. For instance, two excitons with opposite spins can form a molecular bound state, which is called a biexciton. Polaritonic Feshbach resonance corresponds to a scattering process during which the biexcitonic molecular state couples resonantly to the state of two polaritons with anti-parallel spins. The polaritonic Feshbach resonance has been demonstrated on the lower polariton branch [9]. It allows the modification of lower-polariton self-interactions. The interaction between two anti-parallel spin lower-polaritons changes sign abruptly, from attractive to repulsive, when tuning the energy of the two polaritons across the resonance with the biexciton bound state.

The exciton-exciton interaction not only introduces polariton self-interactions but also cross polariton interactions [28,29]. The scheme for generating entangled photons using polariton inter branch scattering has been theoretically proposed [10,11]. It is predicted that entangled photons can be generated by the decay of a biexciton into two polaritons with opposite momenta and spins. This process is expected to be optimal at exciton-cavity detuning comparable to the biexciton binding energy. This corresponds to the condition for which the energy sum of the lower and upper polaritons with opposite spins equalizes the biexciton energy. Under this condition, a cross Feshbach resonance is expected to be observable. The determination of this cross Feshbach resonance is thus, the main step towards the generation of entangled photon pairs in semiconductor microcavities.

In this work, we demonstrate cross Feshbach resonance of a pairing lower and upper polariton coupling to biexciton in semiconductor microcavity. We use spectrally resolved circularly polarized pump-probe spectroscopy (Methods). The lower polaritons are excited resonantly with a circularly polarized narrow-band pump pulse generating a spinor lower polariton population. The cross interaction between upper and lower polariton is probed with a counter-circular polarized probe pulse. We spectrally probe the energy and intensity of the upper polariton peak by measuring the transmission spectrum of the probe pulse. We control the character of the interaction around the cross Feshbach resonance by tuning the cavity-exciton energy. The cross Feshbach resonance is found at a negative cavity detuning for which the total energy of one lower plus one upper polariton matches the biexciton energy. The dynamics around the cross Feshbach resonance is studied by performing time-resolved experiments.

The character of the polariton interaction is detected through the energy shift of the polariton resonance: a blueshift or a redshift respectively mean, a repulsive or an attractive interaction. In Fig. 1, we compare the transmitted probe spectrum of the upper polariton resonance for different cavity detunings, measured at zero delay between pump and probe pulses, with or without the pump. We can track in particular the changes in the energy shift of the upper polariton in the presence of an anti-parallel spin lower polariton population. A change of energy detuning from negative to positive corresponds to an energy shift of the upper polariton being first a redshift, then switching to a blueshift and recovering a zero value. Notice that the amplitude of the probe signal decreases with detuning. In Fig. 2, we plot the energy shift (Fig. 2a) and the intensity variation of the transmitted signal log(I probe/I pump-probe) (Fig. 2b) as a function of cavity detuning. The signature of the cross Feshbach resonance is clearly evidenced at -0.9 meV detuning through the prompt change of sign of the energy shift, which demonstrates the switching of the nature of the interactions between lower and upper anti-parallel spin polaritons from attractive to repulsive. Correspondingly, the maximum reduction of the signal intensity at this detuning shows the resonant conversion of the upper and lower antiparallel spin polariton pair into a biexciton. This optimum cavity detuning also provides a direct measure of the biexciton binding energy $E_{BXX}$ (see insert in Fig. 2a).

The detuning dependence of the probe beam transmission spectrum was modeled solving the Gross-Pitaevskii equations of motion for the two polariton modes (upper and lower branch) and one Heisenberg equation of motion for the biexciton state and assuming CW optical excitations for both pump and probe [24]. Under this assumption, analytical expressions for the dependence with the energy detuning can be obtained for the energy shift and for the transmitted probe amplitude of the upper polariton (see methods). In Figure 2a, b, we compare the detuning dependence of the experimental results with the dependence given by these expressions for the energy shift and for the absorption change, respectively. The best fit with the theoretical expressions was obtained with a biexciton binding energy of 0.9 meV, a biexciton interaction parameter of 1.2 meVcm and a resonance width of 1 meV. From the quality of these fits, we conclude that the Feshbach resonance is indeed caused by the scattering of a pair of upper and lower polariton modes into the biexciton state.

In Fig. 3, we display the transmission of the probe pulse centered on the upper polariton peak as a function of the pump-probe time delay, for a detuning of -1.2 meV, in the vicinity of the cross Feshbach resonance. We observe an energy shift of the upper polariton peak and a reduction of its amplitude around a zero time delay. The temporal dynamics of the transmitted probe pulse characterizes the cross Feshbach resonance. The dynamics is clearly revealed by the dependence of the energy shift with delay. At a zero delay time, the energy shift reaches its maximum value: this corresponds to the largest scattering rate of an upper polariton with a lower polariton to the biexciton state occurring as expected when the optical pulses have the largest temporal overlap. The energy shift varies more rapidly at negative delays than it does at positive delays. The dynamics of the signal at negative delays is governed by the lifetime of the upper polariton state, generated by the probe pulse while, at positive delays, the dynamics is governed by the lifetime of the lower polariton population generated by the pump. Thus, the faster dynamics approaching zero delay from the negative side directly reveals the shorter lifetime of a population of upper polariton in comparison to that of the lower polaritons. The lifetime of the lower polariton population is determined in part by the scattering rate between a lower and an upper polariton with opposite spins to the biexciton state and, for the other part, by its polariton component and the escape rate of the photon from the microcavity for the lower polariton state [30].

The observation of the cross Feshbach resonance is a necessary step towards the generation of entangled photon pairs in semiconductor microcavities. The scattering strength between a biexciton and two antiparallel spin polaritons determines the efficiency of the generation process of a pair of entangled photons with opposite momenta and spins. At the cavity detuning corresponding to the cross Feshbach resonance, the generation of entangled photon pairs could be enhanced by exciting the lower and upper polariton modes with zero in plane momenta and by measuring the temporal correlation in the emission of the two photons in different polarization basis.

## Methods

The sample under investigation is an III-V GaAs-based microcavity [31]. A single 8 nm $In_{0.04}Ga_{0.96}As$

quantum well is sandwiched between a pair of GaAs/AlAs distributed Bragg-reflectors. The Rabi splitting at zero cavity-exciton detuning is 3.45 meV. In order to investigate the cross Feshbach resonance effect, we perform pump and probe experiment in close degenerate configuration ($k_{probe} \approx k_{pump} = 0$ μm$^{-1}$). The broadband few hundred femtosecond pulses are generated by a Ti:sapphire laser with 80 MHz repetition rate and the pump pulse is spectrally narrowed (to 0.5 meV) passing through a single grating pulse shaper. The pump spot size is larger than the probe size to insure a homogeneous pump excitation. The probe intensity is approximately one tenth of the pump intensity of $2.8 \times 10^{11}$ photons pulse$^{-1}$ cm$^{-2}$. The energy of the pump pulse is adjusted to the energy of the lower polariton resonance. We then generate resonantly a spin-up lower polariton population with a $\sigma^+$ circularly polarized pump pulse. The upper polariton resonance is probed with a $\sigma^-$ probe pulse. The probe spectrum is measured in transmission as a function of the exciton-cavity detuning and also as a function of the delay $\tau$ between pump and probe pulses. $\tau>0$ ($\tau<0$) means that the pump (probe) arrives before the probe (pump) pulse. The cavity spacing layer is wedged allowing us to tune the cavity resonance energy by moving the laser spot over the sample. The measurements are performed on the sample at different detunings and at temperature of 4 K.

**Simulation:**
Based on the model described in Ref. [29], we obtain the detuning dependence of the probe transmission spectrum by solving the coupled Gross-Pitaevskii equations of motion for the upper and lower polariton modes and one Heisenberg equation of motion for the biexciton state. The analytical expression for the dependence with the cavity detuning is obtained for the upper polariton energy shift $\Delta E_{U,\downarrow}$:

$$\Delta E_{U,\downarrow} = g^{+-} X_0^2 |C_0|^2 |\psi_{L,\uparrow}^{pu}|^2 + \text{Re}\left[\frac{g_{Bx}^2 X_0^2 |C_0|^2 |\psi_{L,\uparrow}^{pu}|^2}{\varepsilon_L + \varepsilon_U - \varepsilon_B + i\gamma_B}\right]$$

The $X_0$ and $C_0$ are the Hopfield coefficients. The antiparallel spin ↑,↓ lower-upper polariton background interaction constant is given by $g^{+-} X_0^2 |C_0|^2 = g_{UL}^{+-}$ and $g_{Bx}$ is the coupling strength to biexciton. The lower polariton density generated by the pump is $|\psi_{L,\uparrow}^{pu}|^2 = |C_0|^2 n^{pu}$. $\varepsilon_L$ and $\varepsilon_U$ are the energy of the lower and upper polariton, $\varepsilon_B$ and $\gamma_B$ are the biexciton energy and linewidth.

The $\alpha_2 + i\alpha_2'$ real and imaginary part of the expression above gives respectively the interaction constants responsible for the upper polariton energy shift and absorption:

$$\alpha_2 = g^{+-} X_0^2 |C_0|^2 + g_{Bx}^2 X_0^2 |C_0|^2 \frac{\varepsilon_L + \varepsilon_U - \varepsilon_B}{(\varepsilon_L + \varepsilon_U - \varepsilon_B)^2 + \gamma_B^2}$$

$$\alpha_2' = g_{Bx}^2 X_0^2 |C_0|^2 \frac{\gamma_B}{(\varepsilon_L + \varepsilon_U - \varepsilon_B)^2 + \gamma_B^2}$$

The parameters used in the simulation are: $g^{+-} = -1.2$ meV/n$_0$, $g_{Bx} = 1.2$ meV/$\sqrt{n_0}$, $E_{BXX} = -0.9$ meV, $\gamma_B = 0.5$ meV, $n^{pu} = 0.52$ n$_0$ and n$_0 = 5.6 \times 10^{11}$ photons/pulse/cm$^2$.

[1] Inouye, S. et al., Observation of Feshbach resonances in a Bose-Einstein condensate, *Nature* **392**, 151-154 (1998).
[2] Theis, M. et al., Tunning the scattering length with an optically induced Feshbach resonance, *Phys. Rev. Lett. 93, 123001* (2004)
[3] Chin, C. et al., Feshbach resonances in ultracold gases, *Rev. Mod. Phys.* **82**, 1225-1286 (2010).
[4] Bloch, I., Many-body physics with ultracold gases, *Rev.Mod. Phys.* **80**, 885-964 (2008)
[5] Donley E. A. et al, Dynamics of collapsing and exploding Bose-Einstein condensates, *Nature* **412**, 295-299 (2001).
[6] Greiner, M., et al., Emergency of a molecular Bose-Einstein condensate from Fermi gas, *Nature* **426**, 537-540 (2003).
[7] Lopes, R., et al., Quantum depletion of a homogeneous Bose-Einstein condensate, *Phys. Rev. Lett.*

**119**, 190404 (2017)

[8] Cui, Y., et al., Observation of broad d-wave Feshbach resonances with a triplet structure, *Phys. Rev. Lett.* **119**, 203402 (2017)

[9] Takemura, N., et al., Polaritonic Feshbach resonance, *Nat. Phys.* **10**, 500-504 (2014)

[10] Oka, H., et al., High efficient generation of entangled photons by controlling cavity bipolariton states, *Phys.Rev. Lett.* **100**, 170505 (2008)

[11] Oka, H., Efficient generation of energy-tunable entangled photons in a semiconductor microcavitiy, *Appl. Phys.Lett.* **94**, 111113 (2009)

[12] Weisbuch, C., et al., Observation of coupled excite-photon mode splitting in a semiconductor quantum microcavity, *Phys. Rev. Lett.* **69**, 3314-3317 (1992)

[13] Kasprzak, J., et al., Bose-Einstein condensation of exciton polaritons, *Nature* **443**, 409-414 (2006).

[14] Baboux, L., et al., Bosonic condensation and disorder-induced localization in a flat band, *Phys. Rev. Lett.* **116**, 066402 (2016)

[15] Amo a., et al., Superfluidity of polaritons in semiconductor microcavities, *Nat. Phys.* **5**, 805-810 (2009).

[16] Kohnle, V., et al., From single particle to superfluid excitations in a dissipative polariton gas, *Phys. Rev. Lett.* **106**, 255302 (2011).

[17] Lagoudakis, K. G., et al., Quantized vortices in an exciton-polariton fluid, *Nat. Phys.* **4**, 706-710 (2008).

[18] Jacquin, T., et al., Direct observation of Dirac cones and a flatband in a honeycomb lattice for polaritons,*Phys. Rev. Lett.* **112**, 116402 (2014)

[19] Ohadi, H., et al., Spontaneous spin bifurcations and ferromagnetic phase transitions in a spinor exciton-polariton condensate,*Phys. Rev. X* **5**, 031002 (2015).

[20] Boulier, T. et al. Polariton-generated intensity squeezing in semiconductor micropillars, *Nat. Comm.* **5**, (2014)

[21] Adiyatullin, A. F., et al. Periodic squeezing in a polariton Josephson junction, *Nat. Comm.* **8** (2017).

[22] Vladimirova, M. et al. Polariton-polariton interaction constants in microcavities, *Phys. Rev. B* **82**, 075301 (2010)

[23] Takemura, N. et al. Heterodyne spectroscopy of polariton spinor interactions, *Phys. Rev. B* **90**,195307 (2014)

[24] Takemura, N. et al. Spin anisotropic interactions of lower polaritons in the vicinity of polaritonic Feshbach resonance, *Phys. Rev. B* **95**, 205303 (2017).

[25] Paraïso, T. K., et al., Multistability of a coherent spin ensemble in a semiconductor microcavity. *Nature Materials* **9**, 655-660 (2010).

[26] Cerna, R., et al., Ultrafast tristable spin memory of a coherent polariton gas. *Nature Comm.* 3008 (2013)

[27] Abbaspour, H., et al., Spinor stochastic resonance, *Phys. Rev.B* **91**, 155307 (2015).

[28] Takemura, N., et al., 2D Fourier transform spectroscopy of exciton-polaritons and their interactions, *Phys.Rev. B* **92**, 125415 (2015)

[29] Ouellet-Plamondon, c., et al., spatial multistability induced by cross interactions of confined polariton modes, *Phys. Rev. B* **93**, 085313 (2016)

[30] Takemura, N., et al., Coherent and incoherent aspects of polariton dynamics in semiconductor microcavities, *Phys.Rev. B* **94**, 195301 (2016)

[31] Stanley, R. P., et al., Ultrahigh finesse microcavity with distributed Bragg refletors, *Appl. Phys.Lett.* **65**, 1883-1885 (1994)

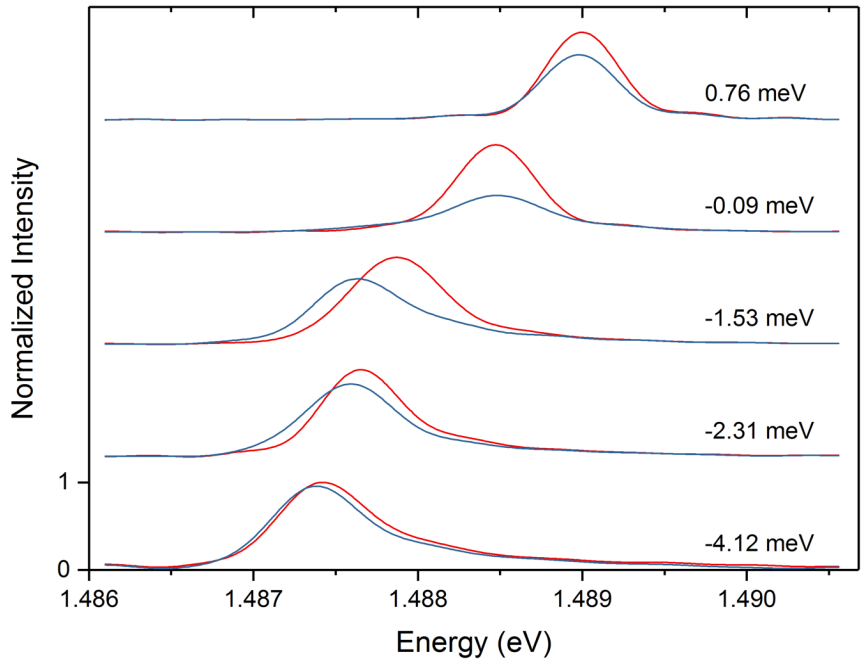

**FIG. 1: Renormalization of the upper polariton peak.** Transmitted probe spectra at upper polariton peak with (blue) and without (red) the presence of the anti-parallel spin lower polariton population for different detunings. The pump-probe spectra are measured at zero time delay. The shift of the energy peak and its amplitude reduction clearly depends on the detuning energy between exciton and cavity (the detuning is given on right hand side of each plot).

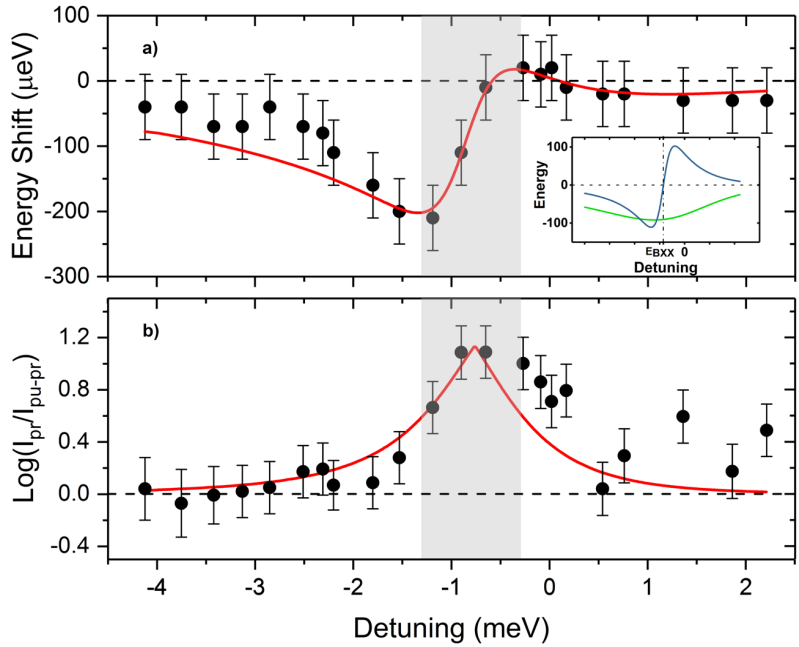

**FIG. 2: Cross Feshbach resonance.** Energy shift **(a)** and intensity variation **(b)** of the upper polariton peak in the presence of the anti-parallel spin lower polariton population as function of cavity detuning. The dots and the solid lines are, respectively, the experimental and numerical simulation results. Fitting parameters are given in simulation section of the methods. The shadow shows the cavity detuning

range, for which the energy sum of the lower and upper polaritons crosses the biexciton energy. In this region the signature of the cross Feshbach resonance is evidenced as an abrupt change of the energy shift and absorption of the upper polaritons. **Insert:** We present the two contributions to the energy shift: from the background interaction (green) and from biexciton scattering (blue). The cross Feshbach resonance occurs for a cavity detuning for which the energy of the antiparallel polariton pair equals the biexciton binding energy $E_{BXX}$.

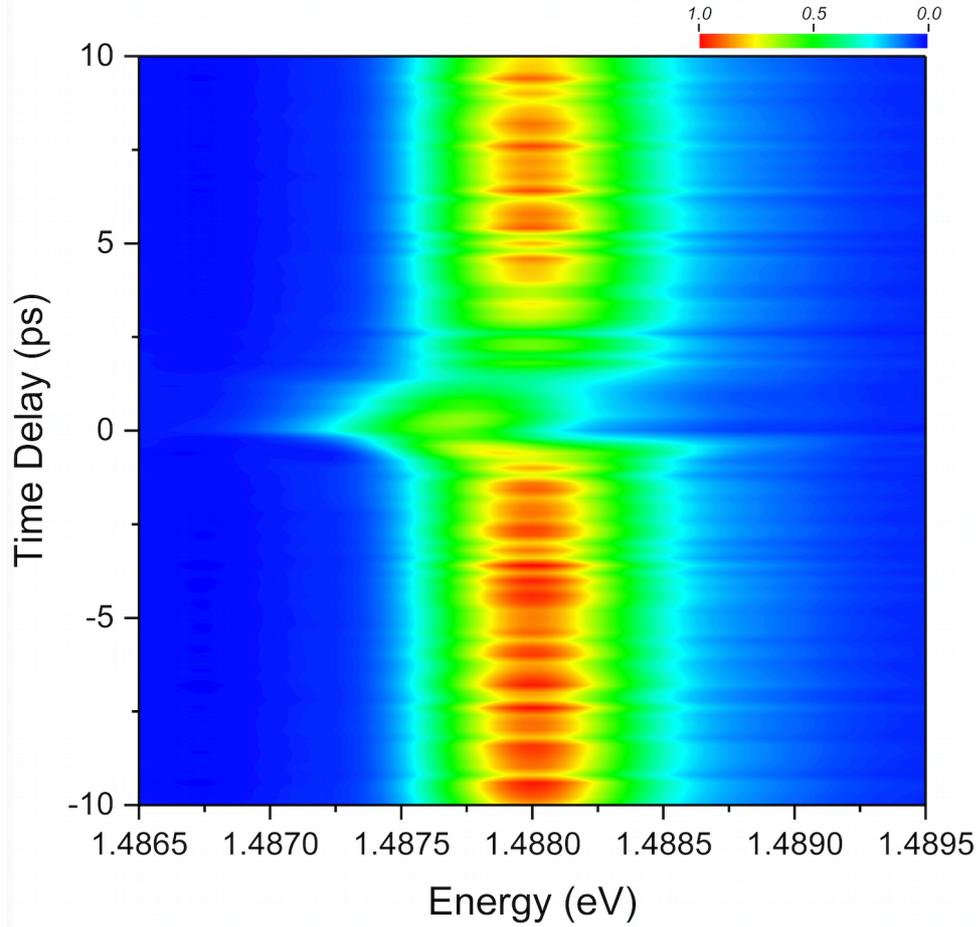

**FIG. 3: Dynamics of the cross Feshbach resonance.** Transmitted probe spectra around the upper polariton peak energy (1.488 eV) as a function of pump-probe delay, measured at -1.19 meV cavity detuning. The largest effect occurs near zero delay. The fast dynamics for positive delays is set by the scattering rate between a lower and an upper polariton with opposite spins to the biexciton.